\documentclass{paper}

\usepackage{amsmath}
\usepackage{todonotes}
\usepackage{tikz}
\usepackage{algorithm}
\usepackage{algpseudocode}

\usepackage{graphicx}
\usepackage{tikz}
\usepackage{amsmath, amsthm, amssymb}
\usepackage{float}
\usepackage[capitalize]{cleveref}
\usepackage[T1]{fontenc}

\newcommand{\etal}{\textit{et~al.}}
\newcommand{\eg}{\textit{e.g.},~}
\newcommand{\ie}{\textit{i.e.},~}

\floatstyle{ruled}
\newfloat{program}{ht}{lop}
\floatname{program}{Algorithm}

\crefname{program}{Algorithm}{Algorithms}
\Crefname{program}{Algorithm}{Algorithms}

\newcommand{\ignore}[1]{}

\makeatletter
\usepackage{etoolbox}
\newcommand*{\@doendeq}{%
  \everypar{{\setbox\z@\lastbox}\everypar{}}%
}
\makeatother

\newcommand{\ISQ}{\mathcal{I}}
\newcommand{\IS}{\eta}
\newcommand{\nullspace}{\phi}
\newcommand{\linkset}{\mathcal{D}}

\newcommand{\mohsen}[1]{}
\newcommand{\localchannel}{\mathcal{C}}
\newcommand{\localtransmission}{\mathcal{T}}

\title{CSMA-SIC: Carrier Sensing with Successive Interference Cancellation}
\author{Mohsen Mollanoori}

\begin{document}
\raggedbottom

\maketitle

\begin{abstract}
Successive interference cancellation~(SIC) is a physical layer
technique that enables the decoders to decode multiple
simultaneously transmitted signals.
The complicated model of SIC requires
careful design of the MAC protocol and
accurate adjustment of transmission parameters.
We propose a new MAC protocol, known as CSMA-SIC,
that employs the multi-packet reception capability of
SIC.
The proposed protocol adjusts the transmission probabilities
to achieve throughput optimality.
\end{abstract}

\section{Introduction}

Traditionally, it is assumed that wireless decoders are
only able to decode one signal at a time, \ie concurrent transmission
of more than one packet results in a collision and
all of the packets need to be re-transmitted.
Based on this underlying assumption, the traditional approach 
to MAC protocol design was to disallow
concurrent transmission of more than one signal.
However, the physical layer of modern wireless networks,
with \textit{multi-packet reception}~(MPR)
capability, is able to decode multiple overlapping
packets transmitted concurrently.
This change to the underlying assumption about the
physical layer calls for a new approach in designing MAC protocols
which encourages concurrent transmissions rather than discouraging them
to take the full advantage of the MPR capability of the physical layer.

\textit{Successive interference cancellation}~(SIC) is a physical layer technique
that employs the structured nature of interference 
to enable MPR in wireless networks.
A decoder with SIC capability, decodes the received signal in
multiple stages.
Since concurrent transmissions happen with SIC,
the received signal is the summation of the 
transmitted signals~(after applying the channel function)
plus the noise signal.
Let $S=S_1+\cdots+S_k+Z$ denote the received signal where
$S_i$ is the signal from the $i$th transmitter,
$k$ is the number of overlapping signals, and
$Z$ is the noise signal.
At each stage, the decoder decodes one of the signals, say $S_j$~(usually
the strongest signal which is not yet decoded).
After the signal $S_j$ is decoded, the analog representation of the
decoded signal is reconstructed and removed
from the summation of the received signals.
In this manner, once the signal of a user is decoded, the summation becomes
free from the interference of that user.
To successfully decode the signals, SIC requires the
transmission parameters~(such as transmission power, rate, and number of 
concurrently transmitted packets) to be carefully controlled.

Using a utility optimization framework proposed by Lin \etal,
the problem of throughput maximization can be decomposed into
congestion control at the transport layer and 
the more difficult job of scheduling at the MAC layer~\cite{lin2006tutorial}.
Considering MPR, because of the more complex interference model,
MAC protocol design becomes even more challenging.

A scheduling algorithm is called throughput optimal
if it stabilizes any arrival rate for which there exists a stabilizing scheduler.
It has been long known~\cite{tassiulas1992stability} that
maximal-weighted scheduling~(MWS) is throughput optimal.
In MWS, at each time slot, a set of non-conflicting links with the maximum
sum of weights are scheduled.
The weight of a link is the length of the queue of that link.
Considering a general interference model, the MWS problem is NP-hard even in
the centralized settings~\cite{sharma2006complexity}.
However, for some special interference models, such as the one hop interference model
discussed by Sharma \etal~\cite{sharma2006complexity}, the problem can be solved in
polynomial time.

To the best of our knowledge, the only
analytical study of scheduling with SIC in multi-hop wireless networks
is by Shabdanov \etal~\cite{shabdanov2012cross}.
The authors propose an optimization framework
to compute the achievable throughput in wireless mesh
networks that employ successive interference cancellation,
superposition coding and dirty-paper coding.
Their framework models each of the problems as a
large scale linear program.
They developed tools based on the column generation method to
compute exact solutions of the models.
However, since the number of variables of the linear programs grows
exponentially with the number of nodes,
the solution is impractical for large networks.
Additionally, since the propose optimization framework requires knowledge of
all nodes and links, the algorithm has to be executed centrally.
The authors also show that superposition coding with SIC capabilities significantly surpasses
other techniques and achieves the maximum theoretical throughput at high power.

Recent studies~\cite{walrand2008allerton, rajagopalan2009network, ni2009distributed, liu2009convergence}
on distributed adaptive \textit{carrier sense multiple access}~(CSMA) protocols offer promising
results both in theory and in simulation. \mohsen{1. I couldn't elaborate more on this. Let's dicuss by phone.}
In CSMA, each node, before beginning the transmission, senses the carrier to
ensure no other conflicting node is transmitting a packet.
Once a node verified that its packet does not conflict with the ongoing
transmissions in the network, it starts the transmission
after waiting for a randomly generated backoff time.

In this paper we propose a distributed scheduling protocol,
inspired by CSMA, which employs SIC at
the physical layer.
The protocol uses the existing RTS/CTS and ACK control packets to alert
the beginning and the end of the transmissions.
Therefore, it can be implemented using little modification to the IEEE 802.11 standard.
We use the framework proposed by Jiang and Walrand~\cite{walrand2008allerton} to analyze the protocol.

\section{System Model} \label{sec:model}
Consider a multi-hop wireless network consisting of $K$ links.
We assume all of the links always have backlog to transmit
since we can simply ignore the links with no backlog.
We also assume that a node cannot transmit and
receive at the same time.

Let $\{1, 2, \dots, n\}$ be the set of the nodes of the network.
Let $g_{ij}$ denote the channel coefficient between nodes $i$ and $j$.
We assume that the channel coefficients between the nodes are
symmetric, \ie $g_{ij}=g_{ji}$ for $i \neq j$.
At any time instance $t$, let $\mathcal{A}_t$ show the set of active transmitters.
The received signal at receiver $i$, denoted by $y_i$, is given by,

\begin{equation}
	y_i = n_i + \sum_{j \in \mathcal{A}_t} g_{ji} S_j
\end{equation}
where $S_j$ denotes the transmitted signal by node $j$ and
$n_i$ is the additive white Gaussian noise signal at node $i$.

For simplicity, it is assumed that all of the transmissions have the same transmission power
and the same rate\footnote{Our proposed protocol can handle multiple rates
	and requires only the control signals be transmitted at a fixed power.}.
It is also assumed that nodes have \textit{successive interference cancellation}
(SIC) capability.
With SIC, decoding is done in multiple stages.
In the first stage, the strongest signal is decoded.
Once a signal is decoded, it is removed from the sum of the signals $y_i$.
In the next stage, the same process is followed for the next strongest signal.
We assume that a signal can be decoded successfully if the 
\textit{signal to interference plus noise ratio}~(SINR) of the signal is more
than a given threshold $\beta$.
More formally, the signal $S_i$ is decodable at the receiver $j$ if,

\begin{equation} \label{eq:snr}
	\frac{|g_{ij} S_i|}{|\sum\limits_{\substack{k \in U_j \\ k \neq i}} g_{kj} S_k + n_j|} \ge \beta
\end{equation}
where $U_j$ is the set of signals that are not decoded yet
and $|S|$ denotes the power of the signal $S$.
Note that, based on our assumptions, a reference signal can be decoded only if
all of the stronger signals can be decoded.

We assume that the signals transmitted by nodes outside the radius $\mathcal{R}$
cannot be decoded and are considered as noise.
The assumption is justified based on the fact that faraway signals
are weak and dominated by near signals.
Furthermore, in a 2D wireless network with
bounded transmission density and path loss exponent~>~2, the expected amount of
cumulative interference from the nodes
beyond radius $\mathcal{R}$
is bounded by a constant\footnote{In 3D, path loss
exponent > 3 is required.}
(see~\cite{brar2006computationally}, \cite{qian2010csma}).
Let $I^{>\mathcal{R}}$ denote this constant.
Using \cref{eq:snr} we define the indicator function $\phi(i, j)$
that specifies whether the signal transmitted from $i$
is decodable at $j$ or not,
\mohsen{2. new indicator function defined}

\begin{equation} \label{eq:snr2}
	\phi(i, j) = \Bigg(
	\frac
	{|g_{ij} S_i|}
	{|\sum\limits_{\substack{k \in U_j^{<\mathcal{R}} \\ k \neq i}} g_{kj} S_k + n_j| + I^{>\mathcal{R}}} \ge \beta
	\Bigg)
\end{equation}
where $U_j^{<\mathcal{R}}$ denotes the set of signals that are transmitted from
within distance $\mathcal{R}$ and are not decoded yet.
Since $I^{>\mathcal{R}}$ is an upper bound for the sum of interference from outside the 
radius $\mathcal{R}$, if $\phi(i, j)$ is true, \cref{eq:snr} holds as well.
This assumption relaxes the requirement of having global knowledge to decide
about the transmissions.
In other words, nodes placed at a distance farther than $\mathcal{R}$ can
decide about their transmissions independently.
This assumption is used in the next section for designing the protocol.

Contention graphs are commonly used to model interference
in CSMA protocol~(\eg see \cite{walrand2008allerton,qian2010csma}).
In a contention graph every link is denoted
by a vertex and there is an edge connecting every pair of conflicting
links.
Contention graphs are only able to represent binary 
relations~(\ie edges between conflicting links), 
while in a network with SIC capability the interference model is too
complex to be representable by binary relations.
For instance, consider the case that $m$ links are active without any conflict
and addition of a new transmission results
in the failure of all of the ongoing transmissions,
which shows an $(m+1)-$ary relation between the links.
Due to this limitation, we use the semantic of \textit{independent sets}
to model the interference in our protocol.

\mohsen{3. updated}
A set of links $\mathcal{L}$ is called an independent set
if for every link in $\mathcal{L}$ such that
$i$ is the transmitter and $j$
is the receiver of the link, $\phi(i, j)$ is true.
Let $\mathbf{v} = \langle v_1,\cdots,v_K \rangle \in \{0,1\}^K$
be the vector representation of independent set $\mathbf{v}$ such that $v_i=1$
if link $i$ is active in independent set $\mathbf v$
and $v_i=0$ otherwise.
For simplicity of exposition we assume that the capacity of
all of the links are equal and normalized to 1,
but our proposed protocol is capable of handling unequal
capacities with little modification~(see \cref{sec:protocol}). \mohsen{4. updated}
Let $\mathcal{IS}$ denote the set of all possible independent sets based on the model above.
The capacity region of the network $\mathcal{C}$ is defined as the set of all rate vectors
that are achievable by time sharing
between the independent sets.
Formally, $\mathcal{C}$ is the affine combination of independent set vectors,

\begin{equation}
	\label{eq:capacity-region}
	\mathcal{C}=\left\{ \sum\limits_{\mathbf{v} \; \in \; \mathcal{IS}} \alpha_\mathbf{v} \mathbf{v}:
		\Big(\sum\limits_{\mathbf{v} \; \in \; \mathcal{IS}} \alpha_\mathbf{v} = 1 \Big) \bigwedge
		\Big(\alpha_\mathbf{v} \ge 0\Big) \;\; \mbox{\Large $\mathsurround0pt\forall$} \mathbf{v}
	\right\}.
\end{equation}

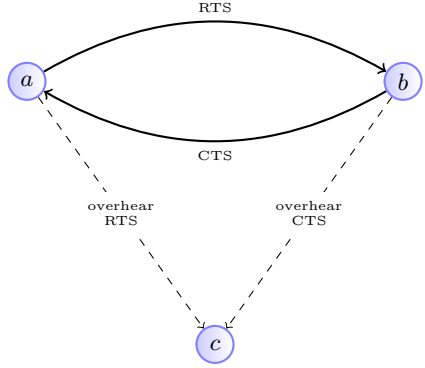
\begin{figure}
	\centering
	\usetikzlibrary{arrows}
	\tikzstyle{NodeStyle}=
	[circle,thick,left color=blue!20,draw=blue!50,minimum size=14pt,inner sep=1pt]
	\begin{tikzpicture}	
	
	\foreach \pos/\name in {{(-2.5,0)/a}, {(2.5,0)/b}, {(0,-3.5)/c}}
	\node[NodeStyle] (\name) at \pos {\small{$\name$}};
	
	\draw[style=->,thick] (a) edge[bend left] node[above]{\tiny{RTS}} (b);
	\draw[style=->,thick] (b) edge[bend left] node[below]{\tiny{CTS}} (a);
	\draw[->,dashed] (a) -- 
		node[fill=white,text width=2cm,align=center]{\tiny{overhear} \\
			\vspace{-0.22cm} \tiny{RTS}} (c);
	\draw[->,dashed] (b) --
		node[fill=white,text width=2cm,align=center]{\tiny{overhear} \\
			\vspace{-0.22cm} \tiny{CTS}} (c);
	\end{tikzpicture}
	\caption{ \label{fig:overhear}
		Node $a$ send an RTS to node $b$.
		Nodes $b$ sends back to $a$ a CTS.
		Node $c$ overhears both of the packets and updates its estimates
		of $g_{ab}$, $g_{ac}$, and $g_{bc}$.
		Note that the CTS packet contains $b$'s estimate of $g_{ab}$.
	}
\end{figure}

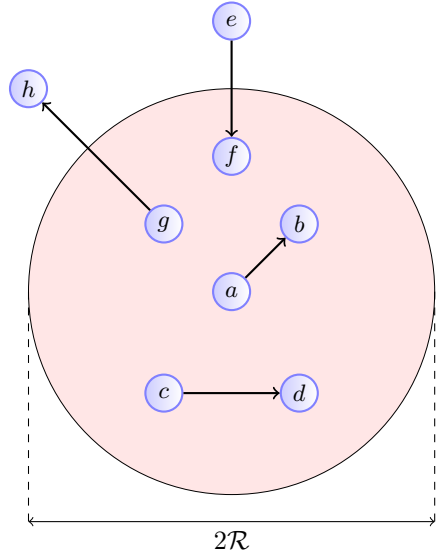
\begin{figure}
	\centering
	\usetikzlibrary{arrows}
	\tikzstyle{NodeStyle}=
	[circle,thick,left color=blue!20,draw=blue!50,minimum size=14pt,inner sep=1pt]
	\begin{tikzpicture}[scale=.9]
	\draw[very thin,fill=red!10] (0, 0) circle (3cm);
	
	\draw[dashed] (-3,0) -- (-3,-3.5);
	\draw[dashed] (+3,0) -- (+3,-3.5);
	\draw[style=<->] (-3,-3.4) -- node[below]{$2\mathcal{R}$} (3,-3.4);
	
	\foreach \pos/\name in
	{{(0,0)/a}, {(1,1)/b}, {(-1,-1.5)/c}, {(1,-1.5)/d}, 
		{(0,4)/e}, {(0,2)/f}, {(-1,1)/g}, {(-3,3)/h}}
	\node[NodeStyle] (\name) at \pos {\small{$\name$}};
	
	\draw[style=->,thick] (a) -- (b);
	\draw[style=->,thick] (c) -- (d);
	\draw[style=->,thick] (e) -- (f);
	\draw[style=->,thick] (g) -- (h);
	\end{tikzpicture}
	\caption{ \label{fig:local-table}
		Demonstration of the local channel coefficient table.
		Node $a$ is placed at the center of the circle with radius $\mathcal{R}$.
		There are four different cases to consider for the entries in $\localchannel_a$:
		(i: link $ab$) Node $a$ transfers the RTS itself and receives the CTS, it updates $g_{ab}$
		(ii: link $cd$) Node $a$ hears both RTS and CTS, it updates $g_{cd}$, $g_{ca}$, and $g_{da}$
		(iii: link $ef$) Node $a$ does not hear the RTS but it hears the CTS, it updates $g_{ef}$ and $g_{fa}$
		(iiii: link $gh$) Node $a$ hears the RTS but not the CTS,
		therefore $g_{gh}$ cannot be estimated by $a$.
		It doesn't cause a problem since $h$ is not in the transmission range of $a$.
		Node $a$ only updates $g_{ga}$.
	}
\end{figure}


We consider an idealized CSMA protocol as is considered by
Jiang and Walrand~\cite{walrand2008allerton}
and Qian \etal~\cite{qian2010csma}.
To ensure that the data packets do not collide with the control packets,
we assume that there is a separate control channel.
As is the case in \cite{walrand2008allerton} and \cite{qian2010csma},
we assume that control signals (\ie RTS/CTS, and ACK),
transmitted within the distance $\mathcal{R}$,
are completed instantaneously without any collision.
We also assume that once a control signal is transmitted, all of the nodes
within radius $\mathcal{R}$ of the transmitter can hear and decode it.
Because of physical limitations such as limited propagation speed and bandwidth
of the signals and limited processing power of the nodes,
the assumptions may be imperfect.
However, we consider these assumptions because
1)~They give us a close approximation of the real world situation.
2)~They make analytical analysis of the protocol possible.

\mohsen{5. two paragraphs merged, some lines removed}
In the traditional CSMA protocol, the transmitters need
to answer the question ``Are any of the neighboring links active?''.
This can simply be answered by sensing the carrier~(see~\cite{jayasuriya2004hidden, xu2002effective}).
But in the CSMA-SIC protocol, the transmitters need to answer 
the question ``Would the currently active neighboring links together with the
new one be able to decode their signals?''.
This is a more complex question that
cannot be answered by only sensing the channel.
The purpose of the control signals in CSMA-SIC are twofold:
to estimate the channel coefficients and
to inform the neighbors about the ongoing transmissions.
It is assumed that channel coefficients
can be estimated using the control signals
(\ie control channel and data channel have reasonably similar channel coefficients).
Nevertheless, if the channel coefficients are known from another source,
we can simply ignore the former purpose.


\section{Continuous Time CSMA Protocol with Successive Interference Cancellation (CSMA-SIC)}
\label{sec:protocol}

Let $\mathcal{A}_i$ denote the area within distance $\mathcal{R}$
of node $i$ and $\mathcal{N}_i$ denote the set of nodes inside $\mathcal{A}_i$.
Each node $i$ keeps two local tables, 
$\localchannel_i$ and $\localtransmission_i$,
and updates them frequently upon hearing a control signal.
Table $\localchannel_i$ keeps the channel coefficient between
all of the nodes in $\mathcal{N}_i$ and
table $\localtransmission_i$ keeps all of the ongoing
transmissions within $\mathcal{A}_i$.
Using $\localchannel_i$ and $\localtransmission_i$,
node $i$ can decide whether starting a new transmission
interrupts the ongoing transmissions~(\ie the transmitted
signal is decodable at the receiver plus the additional interference is not
interrupting the ongoing transmissions of the neighbors) or not.

In our algorithm, each node only needs the channel coefficients
between the nodes in $\mathcal{N}_i$.
Since the control signals are transmitted at a fixed power,
the receivers can estimate the channel coefficient
based on the received signal.
Once a node receives an RTS, it estimates the channel
coefficient between itself and the transmitter of the RTS.
The CTS packet transmitted back contains
the receiver's estimate of the channel coefficient.
Nodes also overhear the RTS/CTS and ACK packets and update
their estimate of channel coefficient between
the transmitter of the control signals and themselves~(see \cref{fig:overhear}).
Since CTS packets contain an estimate of the channel coefficient between
the transmitter and the receiver, all of the nodes that overhear the
CTS, have an estimate of the channel coefficient between the transmitter
and the receiver.
$\localchannel_i$'s are updated frequently upon hearing an RTS/CTS exchange
between the neighbors.
See \cref{fig:local-table} for different possibilities of node placement within
distance $\mathcal{R}$.

\begin{algorithm}
	\caption{\label{alg:csma-sic}Check the feasibility of a new transmission.}
	\small{
	\begin{algorithmic}[1]
		\State{\textbf{global var} \textit{timer}}
		
		\Procedure{Transmit}{$p$}
			\State{$b_i$ = random number exponentially distributed with mean ${R_i}^{-1}$}
			\State{\textit{suspended} $=$ \Call{CheckAllFeasible}{$i$,dst($p$)}}
			\State{start(\textit{timer}, \textit{suspended}, $b_i$),}
			\State{\quad\textbf{on} timeout: send($p$)}
			\State{\quad\textbf{on} $\mathcal{C}_i/\mathcal{T}_i$ update:}

				\State{\quad\quad\textbf{if} \Call{CheckAllFeasible}{$i$,dst($p$)} \textbf{then}}
					\State{\quad\quad\quad resume(\textit{timer})}
				\State{\quad\quad\textbf{else}}
					\State{\quad\quad\quad suspend(\textit{timer})}
				\State{\quad\quad\textbf{end if}}

		\EndProcedure
		
		\Procedure{CheckAllFeasible}{$i$,$j$}\\
			\Comment{Check the feasibility of a new}\\
			\Comment{transmission from $i$ to $j \in \mathcal{N}_i$}
		\State $C = \mathcal{C}_i$
		\State $T = \mathcal{T}_i \cup \{i \rightarrow j\}$

		\For{rx $\mathbf{in}$ receivers of $\mathcal{N}_i$}
			\State{$tx=$ find $t$ s.t. $ \{t \rightarrow rx\} \in T$}
			\If{\Call{CheckFeasible}{$tx,rx,C,T$} == False}
				\State\Return False
			\EndIf
		\EndFor
		\State{\Return{True}}
		\EndProcedure
		
		\Procedure{CheckFeasible}{$tx,rx,C,T$}
			\State{$n_0$ = noise power at node $i$}
			\State{$z$ = percentage of interference canceled by the decoder}
			\State{$\beta$ = minimum SINR required to decode the signal}
			\State{$g_{j}=C_{rx,j}$ $\forall j \in$ transmitters of $\mathcal{N}_i$}
			\State{sort $g_j$'s decreasingly}
			\State{$\nu = \frac{n_0}{P} - g_1 + \sum_j g_j$}\\
				\Comment{$\nu$ is noise plus interference}\\
				\Comment{normalized by transmission power $P$}
			\For{$\mathbf{each}$ $g_j$}
				\If{$\frac{g_j}{\nu} \ge \beta$}
					\If{$j == tx$} \\ \Comment{The desired signal can be decoded}
						\State{\Return True}
					\EndIf
					\State{$\nu = \nu - z g_j$} \Comment{Remove the interference}
				\Else
					\State{\Return False}
				\EndIf
			\EndFor
		\EndProcedure
	\end{algorithmic}
	}
\end{algorithm}

Whenever transmitter $i$ wants to transmit a packet to node $j$,
it generates a random number $b_i$, known as the backoff time,
which is exponentially distributed with mean 
$\frac{1}{R_i}$~(see the next section for more details on $R_i$).
The transmitter starts a timer for $b_i$ units of time.
The timer is paused if the transmission becomes infeasible and
is resumed once the transmission becomes feasible.
Once the timer has timed out, $i$ sends an RTS message to $j$ and
receives back the CTS message.
The RTS/CTS exchange allows the neighbors to update their
local tables.
Then it starts the actual data transmission.
Note that the feasibility of the transmission changes only when
$\localchannel_i$ or $\localtransmission_i$ changes~(\ie upon
hearing a new signal message).
To check the feasibility of the new transmission, each node simulates the
decoding procedure of its neighbors.
See \cref{alg:csma-sic} for the pseudo-code of the algorithm.
Note that we assumed that all of the links have the same capacity
normalized to 1.

\mohsen{7. unequal capacities}
In the case that the links have different capacities,
$\beta$ must be parametrized for each link.
Additionally, control signals must be updated accordingly to contain the required information
to let the transmitters know about the capacities of the neighboring links.

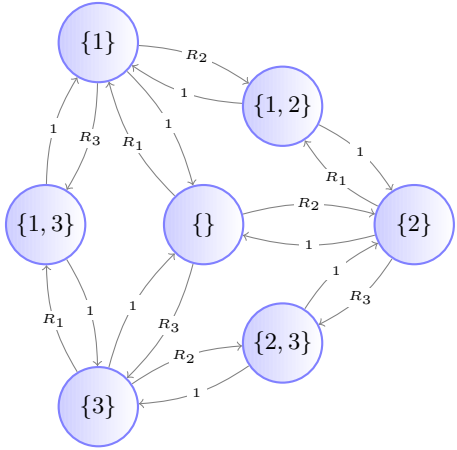
\begin{figure}
	\centering
	\tikzstyle{NodeStyle}=
	[circle,thick,left color=blue!20,draw=blue!50,minimum size=30pt,inner sep=1pt]
	\begin{tikzpicture}[scale=.7]
	
	\foreach \pos/\name/\txt in {{(0,0)/a/$\{\}$}, {(-2,3.46)/b/$\{1\}$}, 
		{(4,0)/c/$\{2\}$}, {(-2,-3.46)/d/$\{3\}$}, {(1.5,2.25)/e/$\{1,2\}$}, 
		{(1.5,-2.25)/f/$\{2,3\}$}, {(-3,0)/g/$\{1,3\}$}}
			\node[NodeStyle] (\name) at \pos {\small{\txt}};

	\foreach \src/\dst/\r in {{a/b/1},{a/c/2},{a/d/3},{b/e/2},{b/g/3},{c/e/1},{c/f/3},{d/f/2},{d/g/1}}
		\draw[->] (\src) edge[bend left=15,thin,gray,text=black] 
			node[fill=white,inner sep=1pt]{\tiny{$R_\r$}} (\dst);
	\foreach \src/\dst/\r in {{a/b/1},{a/c/2},{a/d/3},{b/e/2},{b/g/3},{c/e/1},{c/f/3},{d/f/2},{d/g/1}}
		\draw[->] (\dst) edge[bend left=15,thin,gray,text=black]
			node[fill=white,inner sep=2pt]{\tiny{$1$}} (\src);
	
	\end{tikzpicture}
	\caption{ \label{fig:markov-chain}
		The CSMA-SIC Markov chain for a network with three links in which at most two links
		can transmit simultaneously.
	}
\end{figure}

\section{Achieving Throughput Optimality by Dynamic Adjustment of $R_i$}
In the previous section, we described that the backoff time of link $i$ is
exponentially distributed with mean ${R_i}^{-1}$.
In this section we show that different values of $R_i$ result in different
rate vectors and describe a method to dynamically adjust $R_i$ to achieve
throughput optimality.

Let $K$ be the number of links with backlog in the network.
Assume the backoff time of link $i$ is exponentially distributed with
mean ${\lambda_i}^{-1}$~(\ie link $i$ starts transmission of a 
new packet with rate $\lambda_i$).
Also assume that the duration of the packets at link $i$ is exponentially distributed
with mean $\mu_i$.
Let $\linkset$ denote the set of active links in the network
and let $\linkset+i$ and $\linkset-j$ denote the set of links obtained by adding
the link $i$ to $\linkset$ and removing the link $j$ from $\linkset$ respectively.
Let $\ISQ(\linkset)$ be an indicator function that shows whether the set of links $\linkset$
can be scheduled at the same time or not (considering SIC-enabled physical layer with
the interference model described in \cref{sec:model}).
Define $\IS(\linkset)$ as the set of links that each can be added to $\linkset$ without interrupting
the ongoing transmissions of $\linkset$, \ie $\IS(\linkset) = \{i \; | \; \ISQ(\linkset+i)\}$.

CDMA-SIC protocol can be modeled as a continuous time
Markov chain in which
every independent set~(\ie every $\linkset$ such that
$\ISQ(\linkset)$ holds) is a state.
See \cref{fig:markov-chain} for an example.
There is a transition with rate $\lambda_i$ from the state $\linkset$ to the
state $\linkset+i$ if and only if $i \in \IS(\linkset)$.
There is also a transition with rate $\mu_i$ from the state $\linkset+i$ to the
state $\linkset$ if and only if $i \in \IS(\linkset)$.
The transitions take place only between the states that differ
only in one link.
This follows our assumption that the control signals are
finished instantaneously without any collision.
Therefore, the probability that more than one transition
happen at the same time is zero.

Assuming that the network is in a stable state for a given set of $\lambda_i$ and $\mu_i$, 
the probability of the state $\linkset$ denoted by $Q(\linkset)$
follows the global balance equations~\cite{boorstyn1987throughput},

\begin{align}
	\label{eq:global}
	\left(\sum_{i \in \linkset} \mu_i + \sum_{j \in \IS(\linkset)} \lambda_j \right) Q(\linkset) &=\\
	\sum_{i \in \linkset} \lambda_i Q(\linkset-i) &+ 
	\sum_{j \in \IS(\linkset)} \mu_j Q(\linkset+j) \notag
\end{align}
and the detailed balance equation,

\begin{equation}
	\label{eq:detailed}
	\mu_j Q(\linkset+j)=\lambda_jQ(\linkset)
\end{equation}

Using \cref{eq:global} and \cref{eq:detailed} we have,

\begin{equation}
	\label{eq:steady-state}
	Q(\linkset)=\left( \prod_{i \in \linkset} \frac{\lambda_i}{\mu_i} \right)Q(\nullspace)
\end{equation}
where $\linkset$ is a set of nodes such that $\ISQ(\linkset)$ holds
and $\nullspace$ denotes the null space (\ie the state in which no node is transmitting).
To have a steady state, it is required to have $Q(\nullspace) > 0$
\ie $\nullspace$ is positive recurrent.
Using the equality $\sum_\linkset Q(\linkset)=1$ we can compute $Q(\nullspace)$.
More details on the described modeling technique can be found in~\cite{boorstyn1987throughput}.
Let $\lambda_i=R_i=e^{r_i}$.
To simplify the analysis we assume that the duration of the packets are exponentially distributed
with mean 1, \ie $\mu_i=1$.
We can rewrite \cref{eq:steady-state} as,

\begin{equation}
	\label{eq:steady-state2}
	Q(\linkset)=\exp\left( {\sum_{i \in \linkset}r_i} \right)Q(\nullspace)
\end{equation}
where,

\begin{equation}
	\label{eq:null-state-prob}
	Q(\nullspace)=\frac{1}{\sum_\linkset \exp\left( {\sum_{i \in \linkset}r_i} \right)}.
\end{equation}

From \cref{eq:steady-state2} and \cref{eq:null-state-prob} it is apparent that
any given set of $r_i$ results in a different distribution for $Q(.)$.
Let $\mathbf{r}=\langle r_1, \cdots, r_m \rangle$, where $m$ denotes the number
of independent sets.
By $Q_\mathbf{r}(\linkset)$ we denote the steady state probability
of the state $\linkset$ given $\mathbf{r}$.
The expected throughput of the link $i$ is then given by,

\begin{equation} \label{eq:expected-throughput}
	\tau_i(\mathbf{r})=\sum_{\linkset|i \in \linkset} Q_\mathbf{r}(\linkset)
\end{equation}

Let $\mathbf{x} \in \mathcal{C}$ be a given rate vector inside the capacity
region~(see \cref{eq:capacity-region}).
Jiang and Walrand~\cite{walrand2008allerton}, show that using a simple distributed
gradient decent algorithm, which updates $\mathbf{r}$ during the time,
the network will reach the steady state that supports $\mathbf{x}$.
That is, $x_i \ge \tau_i(\mathbf{r})\; \forall i$.
The algorithm updates the value of $\mathbf{r}$ at time
$t_1 < t_2 < t_3 < \cdots$.
Let $\mathbf{r}(i)$ be the value of $\mathbf{r}$ at time $t_i$.
Set $\mathbf{r}(0)=\mathbf{0}$ and update $\mathbf{r}(i)$ at time
$t_i$ for $i=1,2,3,\cdots$ using the following equality,

\begin{equation}
	\label{eq:gradient-decent}
	r_k(i)=[r_k(i-1)+\alpha(i) (\lambda'_k(i)-\tau'_k(i))]_+
\end{equation}
where $\alpha(i)$ is a small step size and $\lambda'_k(i)$ and $\tau'_k(i))$
are the empirical values of the arrival and service rates during time $t_{i-1}$
to $t_i$.
The $[\cdot]_+$ operator sets $r_k(i)=0$ when it becomes negative.
\Cref{eq:gradient-decent} actually increases~(decreases) $r_k$, and as a result
the transmission rate, in the case the 
arrival rate is more~(less) than the service rate during the time $t_{i-1}$
to $t_i$.

\section{Conclusion}

SIC changes the basic properties of physical layer of
wireless networks and as a result it changes the underlying assumptions
of MAC protocol design.
Specially designed MAC protocol is required to fully utilize the 
multi-packet reception capability of SIC.
CSMA-SIC is a MAC protocol inspired by CSMA and employs SIC at
the physical layer.
The proposed protocol uses the existing RTS/CTS and ACK control packets
to coordinate the transmissions.
Thus it incurs little overhead.
The protocol is throughput optimal in the sense that it
stabilizes any arrival rate for
which there exists a stabilizing scheduler.

\section*{Acknowledgment}
Discussions with Dr.~Majid Ghaderi shaped the initial idea of this paper.
Dr.~Ghaderi also provided the author with some useful references related to
the topic, including referring the author to Javad Ghaderi.
The author appreciates his time and effort.

The present paper benefited form the inputs of Dr.~Javad~Ghaderi who provided
valuable assistant for understanding the topic.
He verified the proposed protocol and mentioned a few issues
in the initial draft of the paper to be fixed.

Dr.~Lisa~Higham had a key role in improving the presentation of the paper.
The author would like to thank Dr.~Higham for her valuable comments and advice
on writing the paper.

\pagebreak
\bibliography{refs}
\bibliographystyle{plain}

\end{document}